\documentclass[accepted]{article}

\usepackage{microtype}
\usepackage{graphicx}
\usepackage{subfigure}
\usepackage{booktabs} 

\usepackage{hyperref}

\usepackage{icml2022}

\usepackage{amsmath}
\usepackage{amssymb}
\usepackage{mathtools}
\usepackage{amsthm}

\usepackage[capitalize,noabbrev]{cleveref}

\theoremstyle{plain}

\theoremstyle{definition}

\theoremstyle{remark}

\usepackage[textsize=tiny]{todonotes}

\usepackage[acronym, shortcuts, nohypertypes={acronym}]{glossaries}
\newacronym{CCC}{CCC}{concordance correlation coefficient}
\newacronym{CNN}{CNN}{convolutional neural network}
\newacronym{EXVO}{\textsc{EXVO}}{ICML Expressive Vocalisations Workshop \& Competition 2022}
\newacronym{FFNN}{FFNN}{feed-forward neural network}
\newacronym{SER}{SER}{speech emotion recognition}
\newacronym{SGD}{SGD}{stochastic gradient descent}

\newcommand{\wrt}{w.\,r.\,t.\ }

\icmltitlerunning{Redundancy Reduction Twins Network: A Training Framework}

\begin{document}

\twocolumn[
\icmltitle{Redundancy Reduction Twins Network: \\A Training framework for Multi-output Emotion Regression}



\icmlsetsymbol{equal}{*}

\begin{icmlauthorlist}
\icmlauthor{Xin Jing}{yyy}
\icmlauthor{Meishu Song}{yyy,zzz}
\icmlauthor{Andreas Triantafyllopoulos}{yyy}
\icmlauthor{Zijiang Yang}{yyy}
\icmlauthor{Bj\"{o}rn W. Schuller}{yyy,xxx}
\end{icmlauthorlist}

\icmlaffiliation{yyy}{EIHW, University of Augsburg, Augsburg, Germany}
\icmlaffiliation{zzz}{Educational Physiology Laboratory, University of Tokyo, Japan}
\icmlaffiliation{xxx}{GLAM, Imperial College, London, United Kingdom}

\icmlcorrespondingauthor{Xin Jing}{xin.jing@uni-a.de}

\icmlkeywords{ExVo, Deep Nerual Network}

\vskip 0.3in
]



\printAffiliationsAndNotice{}  

\begin{abstract}
In this paper, we propose the Redundancy Reduction Twins Network (RRTN), a redundancy reduction training framework that minimizes redundancy by measuring the cross-correlation matrix between the outputs of the same network fed with distorted versions of a sample and bringing it as close to the identity matrix as possible. RRTN also applies a new loss function, the Barlow Twins loss function, to help maximize the similarity of representations obtained from different distorted versions of a sample. However, as the distribution of losses can cause performance fluctuations in the network, we also propose the use of a Restrained Uncertainty Weight Loss~(RUWL)~ or joint training to identify the best weights for the loss function. Our best approach on CNN14 with the proposed methodology obtains a CCC over emotion regression of $.678$ on the ExVo Multi-task dev set, a $4.8\%$ increase over a vanilla CNN14 CCC of $.647$, which achieves a significant difference at the 95\% confidence interval (2-tailed).
\end{abstract}

\section{Introduction}
\label{sec:intro}
The expression of emotions in speech sounds and the corresponding ability to perceive such emotions are both fundamental aspects of human communication \cite{mauss2009measures, latif2021survey}. Understanding the vocal bursts and non-lingistic vocalizations, which are central to the expression of emotion \cite{scherer2017computational}, has been overlooked in the field of computer audition. In this aspect, the \ac{EXVO} \cite{BairdExVo2022} presents an excellent opportunity to take advantage of recent advances in affective computing to study the automatic recognition of emotions through vocalizations.

In supervised learning, data augmentation \cite{Yang20ARC, Triantafyllopoulos21DSC} is typically applied to enlarge the data set in order to prevent overfitting and improve generalization of the models. 
Self-supervised Learning~(SSL) adopts a different means for the same goal by attempting to learn generic representations of massive data without human annotations \cite{liu2022audio}. 
Since there is no label in the training upstream, recent advances in SSL suggest that it is possible to obtain competitive results compared to supervised learning by creating pseudo-labels for training data through data augmentation \cite{zbontar2021barlow}, or by defining `positive' and `negative' pairings \cite{chen2020simclr, baevski2020wav2vec}. Redundancy-reduction is a principle first proposed in neuroscience by H.\ B.\ Barlow \cite{barlow1961possible} in 1961. The principle hypothesizes that the goal of sensory processing is to recode highly redundant sensory inputs into a factorial code with statistically independent components and successfully explain the organization of the visual system \cite{barlow2001redundancy}. It has led to several different works on supervised and unsupervised learning \cite{balle2016end, deco1997non}. Following that principle, this work is mainly inspired by \cite{zbontar2021barlow, barlow1961possible, goyal2019scaling}, where the original data and its augmented version are fed into the same network and their outputs should have large similarity with slight difference. 

Several works have shown that the combination of different loss functions is beneficial when training deep neural networks \cite{bentaieb2017uncertainty, babaeizadeh2018adjustable}, as this allows for incorporating heterogeneous information. For a multi-losses function training, it is a common approach to balance the different properties by minimizing a loss function that is the weighted sum of the different losses \cite{dosovitskiy2019you, liebel2018auxiliary}.  

The key contributions of our work are two-fold:
\begin{enumerate}
    \item We present a simple, but effective redundancy reduction method that generates two distorted perspectives for every sample in the batch and then reduces redundancy in the feature map by operating on the cross-correlation matrix of the resulting embeddings.
    
    \item We apply a novel weighted loss strategy through weight updates and restrain the sum of losses to find the best weight for each loss.
\end{enumerate}


The remainder of our paper is organized as follows. We present our Redundancy Reduction Twins network architecture and Restrained Uncertainty Weight Loss strategy in Section \ref{sec:meth}. In Section \ref{sec:exp} and Section \ref{sec:res}, we describe our experimental setting and results respectively. Finally, we conclude with a brief summary and outline the future directions.

\section{Methodology}
In this section, we propose the Redundancy Reduction Twins Network~(RRTN), a redundancy reduction training framework which applies a pair of identical networks to measure the similarity of distorted versions of a sample and makes the cross-correlation matrix between them as close to the identity matrix as possible. In addition, we also apply a loss strategy named Restrained Uncertainty Weight Loss (RUWL), which dynamically adjusts the weights of different losses and keeps the sum of the weights close to a constant.

\begin{figure*}[t!]
\centering
\centerline{\includegraphics[width=0.77\linewidth]{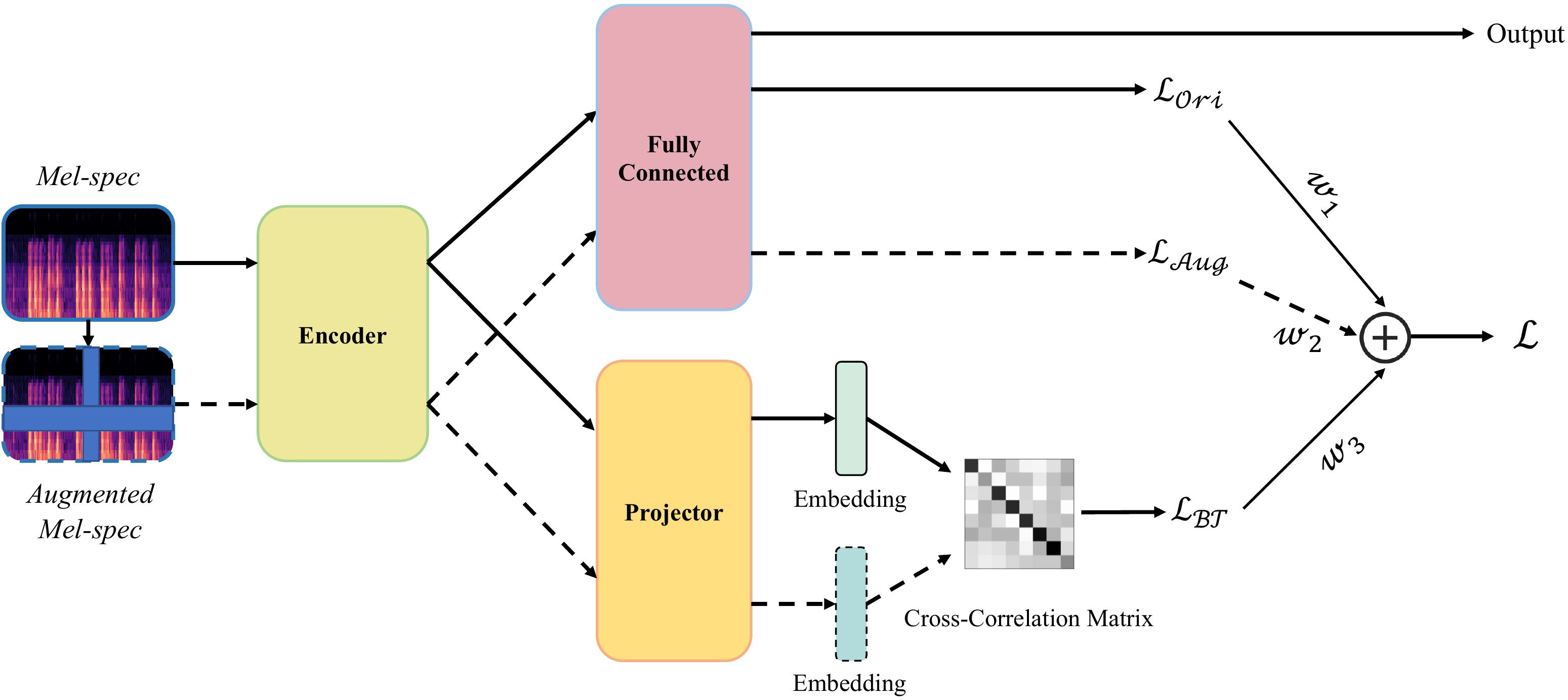}}
\caption{The proposed Redundancy Reduction Twins Network requires distorted versions of a batch of samples and tries to measure the similarity of two embeddings using the correlation matrix. It is designed to mitigate redundancy by trying to equate the diagonal elements of the cross-correlation matrix to 1. The introduced Restrained Uncertainty Weight Loss will dynamically adjust the loss weight while keeping the sum of the weights approaching the upper limit.}
\label{fig:network}
\vskip -0.2in
\end{figure*}
\begin{table}
\centering
\caption{An overview of the Hume-VB dataset. The information of train and validation sets is publicly accessible while the test set is blinded.}
\label{tab:dataset}
\begin{tabular}{l|rrr}
\toprule
\multicolumn{1}{c|}{} & \multicolumn{1}{c}{Train} & Validation & $\sum$ \\ 
\midrule
Minutes & 739 & 726 & 1465 \\
No. & 19990 & 19396 &  59201\\
\midrule
Speakers & 571 & 568 &  1139\\
USA & 206 & 206 &  -\\ 
China & 79 & 79 &  -\\
South Africa & 244 & 244 & - \\
Venezuela & 44 & 44 & -\\
\bottomrule
\end{tabular}
\end{table}

\label{sec:meth}
\subsection{Redundancy Reduction Twins Network}

\autoref{fig:network} shows the architecture of our proposed Redundancy Reduction Twins Network~(RRTN)~training framework. We employ the original input sample and its augmented form as the network's input data. Given \autoref{fig:network}, we refer to the output of its encoder as its \textbf{`representations'} and the output of the projector as its \textbf{`embeddings'}. The representations are used for the regression task and the embeddings are fed to the Barlow Twins loss function. 

For our best approach, which we will discuss in the \autoref{sec:res}, the encoder consists of a CNN14 network\cite{Kong19-PANNS} (without the classification layer, and 2048 output units). The following projector network is different from the original Barlow Twins network\cite{zbontar2021barlow}, which contains 3 linear layers, each with 8192 units. 
Our projector includes only 1 linear layer which has 2048 units, as our method does not benefit from very high embedding. From the \autoref{equ:btl}, the Barlow Twins loss is the sum of elements in a matrix, and when the dimension of embedding rises, it can cause a loss value that is thousands of times larger than the regression loss, which makes it hard to jointly optimize.

\subsection{Loss functions}
\textbf{Concordance Correlation Coefficient Loss~(CCC Loss):} is the concordance between prediction~($y$)~and the ground truth~($x$). This statistic quantifies the agreement between these two measures of the same variable. The CCC loss is defined as
\begin{equation}
 \label{equ:ccc}
  \mathcal{L}_{ccc} = 1 - \frac{2\sigma_{xy}}{\sigma^2_{x}+\sigma^2_{y}+(\mu_x-\mu_y)^2},
\end{equation}

where $\mu_x=\mathbb{E}(\mathbf{x}), \mu_y=\mathbb{E}(\mathbf{y}), \sigma_x^2=var(\mathbf{x}), \sigma_y^2=var(\mathbf{y})$, and $\sigma_{xy}^2=cov(\mathbf{x}\mathbf{y})$. In the considered  emotion recognition task, ten emotion labels are provided, and we use the CCC defined and provided by the ExVo challenge \cite{BairdExVo2022}, which is the mean for each emotion dimension: $\hat C = \sum\nolimits_{i=1}^{10} C_i/10$.

\textbf{Barlow Twins Loss~(BT Loss):} is designed to maximize similarity between the embedding vectors while reducing the redundancy between their components. The BT loss is defined as
\begin{equation}
 \label{equ:btl}
  \mathcal{L}_{BT} \triangleq \sum_{i}(1-\mathcal{C}_{ii})^2 + \lambda \sum_{i}\sum_{j\neq i}\mathcal{C}_{ij}^2,
\end{equation}

in which $\lambda$ is a postive constant that weighs the importance of the first and second term in the loss function. In this work, we set $\lambda=0.001$. $\mathcal{C}$ is the cross-correlation matrix, which is 
\begin{equation}
 \label{equ:ccr}
    \mathcal{C}_{ii} \triangleq \frac{\sum\nolimits_{b}z_{b,i}^A z_{b,j}^B}{\sqrt{\sum\nolimits_{b}(z_{b,i}^A)}\sqrt{\sum\nolimits_{b}(z_{b,j}^B)}},
\end{equation}

where $b$ is the batch smaples index and $i,j$ is the vector dimension of the network output. Both terms in the definition serve a distinct purpose. The first term in \autoref{equ:ccr} tries to equate the diagonal elements of the cross-correlation matrix to 1, which makes the embedding invariant to the distortions applied, while the second term, which is called the redundancy reduction term, decorrelates the different vector components of the embedding by trying to equate the off-diagonal elements of the cross-correlation matrix to 0.

\subsection{Restrained Uncertainty Weight Loss~(RUWL)}
From the \autoref{equ:ccc}, the upper limit of the CCC loss is 1. Meanwhile, as \autoref{equ:btl} shows, the BT loss is a sum of matrices. For a high-dimension embedding in our work, which is 2048, the value of the BT loss could be much higher than the CCC loss. 

As shown in \autoref{fig:network}, in order to optimise the network with three losses, we add them by different weights, which is
\begin{equation}
 \label{equ:loss_sum}
    \mathcal{L}_{sum} = w_1 \times \mathcal{L}_{ccc} + w_2 \times \mathcal{L}_{ccc_a} + w_3 \times \mathcal{L}_{BT},
\end{equation}

where $w_n$ is the weight of different losses and $\mathcal{L}_{ccc_a}$ is the loss of the distorted input sample. 

Since each of the contributing single loss functions may behave differently, appropriately  weighting each loss is essential.
We apply the weighted loss function described in \cite{liebel2018auxiliary} to automatically tune $w_n$. Furthermore, to prevent the network from constantly shrinking the weights to reduce the loss without learning anything about the task, we propose a Restrained Uncertainty Weight Loss for the $\mathcal{L}_w$ to update, which is

\begin{equation}
 \label{equ:loss_w}
    \mathcal{L}_{w} = |2 - |w_1| - |c_{w_2}| - |c_{w_3}||.
\end{equation}

where $c_{ccc}$
The final loss for our RRTN framework is defined as
\begin{equation}
 \label{equ:loss_a}
    \mathcal{L}_{\tau + 1} = \mathcal{L}_{w} + \sum_{\tau \in I}\frac{1}{\lambda_{\tau} \times c_{\tau}^2} \times \mathcal{L}_{\tau} + ln(1+c_{\tau}^2),
\end{equation}

where $c_{\tau}$ a is the parameter that dynamically updates the weights as the network is updated, and $\lambda_{\tau}$ is a constant for each loss. Based on our prior knowledge, in our work, we set $[\lambda_{\mathcal{L}_{ccc}}, \lambda_{\mathcal{L}_{ccc_a}}, \lambda_{\mathcal{L}_{BT}}]=[1, 1, 10^{-8}]$, while initialing the parameters $[c_{\mathcal{L}_{w_1}}, c_{\mathcal{L}_{w_2}}, c_{\mathcal{L}_{w_3}}]=[1, 1, 0.01]$.

\begin{table*}[t]
\centering
\caption{Overall results for our proposed RRTN framework by Concordance Correlation Coefficient~(CCC, \wrt $\hat C$). The parameters of the networks are also presented. Relative Local Gain indicates the improvement on each network itself. The Relative Global Gain, on the other hand, refers to the improvement over the ResNet 18 baseline. The best result of our experiments is marked in \textbf{bold}, and the second best one is \underline{underlined}.}
\label{tab:results}
\begin{tabular}{l|rrrr} 
\toprule
\multicolumn{1}{c|}{Encoder} & \multicolumn{1}{c}{\# Param(M)} & $\hat C$ & Relative Local Gain~(\%) & Relative Global Gain~(\%) \\ 
\midrule
ResNet 18~(baseline) & 11.18 & 0.613 & - & - \\
ResNet 18 $w/o$ RUWL & 12.23 & 0.628 & 2.4 & 2.4 \\
ResNet 18 $w$ RUWL & 12.23 & 0.634 & 3.6 & 3.6 \\ 
\midrule
CNN10 & 4.95 & 0.655 & - & 6.9 \\
CNN10 RRTN $w/o$ RUWL & 6.00 & 0.667 & 1.8 & 8.8 \\
CNN10 RRTN $w$ RUWL & 6.00 & 0.668 & 2.0 & 9.0 \\ 
\midrule
CNN14 & 79.69 & 0.647 & - & 5.5 \\
CNN14 RRTN $w/o$ RUWL & 83.95 & \underline{0.674} & 4.2 & 10.0 \\
CNN14 RRTN $w$ RUWL & 83.95 & \textbf{0.678} & 4.8 & 10.6 \\
\bottomrule
\end{tabular}
\end{table*}

\section{Experiments}
\label{sec:exp}
\subsection{Dataset}
The Hume Vocal Bursts dataset~(HUME-VB)\cite{Cowen2022HumeVB}\footnote{We use the original release of processed .wav files, not the follow-up release of unprocessed ones.} is a large-scale self-recording emotional non-linguistic vocalizations~(vocal bursts)~ dataset, which contains totally 2,270 minutes of audio data from 1,702 speakers, aged from 20 to 39\,years old. 

In the database, each vocal burst has been labeled in terms of the intensity of ten different expressed emotions. Each emotion's intensity ratings were normalized to a range of [0:1]. The audio files were normalized to -3 decibels, and converted to 16\,kHz, 16\,bit, mono. Finally, the official data have been split into three equal size sets: training, validation, and test sets. There are ten emotion labels in the dataset: \textit{Amusement, Awe, Awkwardness, Distress, Excitement, Fear, Horror, Sadness, Surprise, and Triumph}.

\subsection{Experiment settings}
As input features, we use 64-dimensional log Mel-Spectrograms with a 32\,ms window length and a 10\,ms frame shift. To unify the shape of the input data consistently, we crop/pad the time dimension to 250. Finally, the size of the features for the RRTN is $[batch\_size, 1, 250, 64]$. Each input sample is augmented by SpecAugment\cite{Park_spec_2019} to produce the distorted views shown in \autoref{fig:network}. We set SpecAugment with 64 times drop width, 2 time stripes, 8 frequency drop width, and 2 frequency stripes.

Throughout our experiments, we use the AdamW optimiser with an epsilon value of $10^{-8}$ and weight decay of $0.01$. We train for 60 epochs with a mini-batch size of $128$ and a learning rate of $0.001$. All models were developed on Pytorch 1.8.1 and trained on a single Nvidia RTX 3090 GPU.

\section{Results and Discussion}
\label{sec:res}
Our experimental results obtained in the multi-output regression of the vocal emotion task are presented in \autoref{tab:results}, and we follow the original network implementation described in \cite{he2016deep, Kong19-PANNS} to build ResNet 18 and CNN 10/CNN 14. 

As shown in \autoref{tab:results}, in our experiments, we set the ResNet 18 result as a global baseline. We fit each network into the RRTN framework; for the global gain, our method on CNN 14 obtains the highest improvement with $\hat C$ of 0.678, which is a 10.6\,\% increase. Further experiments on the RUWL reveal that it is a simple and effective practice to apply weighted loss to improve the prediction ability of the network.

By applying the RRTN training framework, there is always a gain in the $\hat C$ when compared to the outcomes of the original network, which we dub local gain. However, we find that the improvement in the results obtained by using RRTN with various encoders is quite variable. CNN 14 improves by up to 4.8\,\%, while CNN 10 only improves by up to 2\,\%. CNN 10 and CNN 14 share the same network structure, consisting of 4 and 6 convolutional blocks, respectively. Each convolutional block consists of 2 convolutional layers with a kernel size of $3\times3$.
The goal of the RRTN is to improve the generalizability of the encoder's features by maximizing the similarity and eliminating the redundant information between pairs of samples. The deeper and wider network is more conducive to the joint optimization strategy. 
In our work, RRTN only introduces a linear layer as the projector, which does not significantly increase the parameters of the network, but may lack in representational power. 
Further experiments should be conducted to find the optimal number of linear layers in the projector.
 
Overall, it seems that the most gain to the performance is from the introduction of the RRTN framework, while RUWL adds a marginal gain over that. This means minimizing redundant information in the features is the most important contribution of our proposal, while the automatic weighting of the different losses is not as critical.

\section{Conclusion}

In this work, we proposed the Redundancy Reduction Twins Network (RRTN), which is a light and effective training framework. The cross-correlation matrix was used to quantify the similarity between the outputs of the same network fed with distorted versions of a sample, and it is made as close to the identity matrix as feasible to maximize similarity. Since the final loss is the sum of multiple weighted losses, we applied the restrained uncertainty weight loss for joint training to find the best weight for each loss. Our experimental results demonstrate the validity of our proposed methodologies. 

Moreover, we believe that further refinements of the proposed loss function and framework could lead to more efficient solutions and better performance.
\label{sec:con}


\section{Acknowledgements}
\label{sec:Acknowledgment}
This work was funded by the China Scholarship Council~(CSC), Grant \#\,202006290013.
\newpage
\bibliography{ref.bib}

\begin{thebibliography}{23}
\providecommand{\natexlab}[1]{#1}
\providecommand{\url}[1]{\texttt{#1}}
\expandafter\ifx\csname urlstyle\endcsname\relax
  \providecommand{\doi}[1]{doi: #1}\else
  \providecommand{\doi}{doi: \begingroup \urlstyle{rm}\Url}\fi

\bibitem[Alan et~al.(2022)Alan, Alice, Panagiotis, Michael, Lauren, Jeff, and
  Jacob]{Cowen2022HumeVB}
Alan, C., Alice, B., Panagiotis, T., Michael, O., Lauren, K., Jeff, B., and
  Jacob, M.
\newblock The hume vocal burst competition dataset (h-vb) | raw data [exvo:
  updated 02.28.22] [data set].
\newblock \emph{Zenodo}, 2022.
\newblock \doi{https://doi.org/10.5281/zenodo.6308780}.

\bibitem[Babaeizadeh \& Ghiasi(2018)Babaeizadeh and
  Ghiasi]{babaeizadeh2018adjustable}
Babaeizadeh, M. and Ghiasi, G.
\newblock Adjustable real-time style transfer.
\newblock \emph{arXiv preprint arXiv:1811.08560}, 2018.

\bibitem[Baevski et~al.(2020)Baevski, Zhou, Mohamed, and
  Auli]{baevski2020wav2vec}
Baevski, A., Zhou, Y., Mohamed, A., and Auli, M.
\newblock wav2vec 2.0: A framework for self-supervised learning of speech
  representations.
\newblock \emph{Advances in Neural Information Processing Systems},
  33:\penalty0 12449--12460, 2020.

\bibitem[Baird et~al.(2022)Baird, Tzirakis, Gidel, Jiralerspong, Muller,
  Mathewson, Schuller, Cambria, Keltner, and Cowen]{BairdExVo2022}
Baird, A., Tzirakis, P., Gidel, G., Jiralerspong, M., Muller, E.~B., Mathewson,
  K., Schuller, B., Cambria, E., Keltner, D., and Cowen, A.
\newblock The icml 2022 expressive vocalizations workshop and competition:
  Recognizing, generating, and personalizing vocal bursts, 2022.

\bibitem[Ball{\'e} et~al.(2017)Ball{\'e}, Laparra, and
  Simoncelli]{balle2016end}
Ball{\'e}, J., Laparra, V., and Simoncelli, E.~P.
\newblock End-to-end optimized image compression.
\newblock In \emph{Proceedings of International conference on learning
  representations}, Toulonk, France, 2017.

\bibitem[Barlow(2001)]{barlow2001redundancy}
Barlow, H.
\newblock Redundancy reduction revisited.
\newblock \emph{Network: computation in neural systems}, 12\penalty0
  (3):\penalty0 241, 2001.

\bibitem[Barlow et~al.(1961)]{barlow1961possible}
Barlow, H.~B. et~al.
\newblock Possible principles underlying the transformation of sensory
  messages.
\newblock \emph{Sensory communication}, 1\penalty0 (01), 1961.

\bibitem[BenTaieb \& Hamarneh(2017)BenTaieb and
  Hamarneh]{bentaieb2017uncertainty}
BenTaieb, A. and Hamarneh, G.
\newblock Uncertainty driven multi-loss fully convolutional networks for
  histopathology.
\newblock In \emph{Intravascular Imaging and Computer Assisted Stenting, and
  Large-Scale Annotation of Biomedical Data and Expert Label Synthesis}, pp.\
  155--163. Springer, 2017.

\bibitem[Chen et~al.(2020)Chen, Kornblith, Norouzi, and Hinton]{chen2020simclr}
Chen, T., Kornblith, S., Norouzi, M., and Hinton, G.
\newblock A simple framework for contrastive learning of visual
  representations.
\newblock In \emph{Proceedings of International conference on machine
  learning}, pp.\  1597--1607, virtual, 2020.

\bibitem[Deco \& Parra(1997)Deco and Parra]{deco1997non}
Deco, G. and Parra, L.
\newblock Non-linear feature extraction by redundancy reduction in an
  unsupervised stochastic neural network.
\newblock \emph{Neural networks}, 10\penalty0 (4):\penalty0 683--691, 1997.

\bibitem[Dosovitskiy \& Djolonga(2019)Dosovitskiy and
  Djolonga]{dosovitskiy2019you}
Dosovitskiy, A. and Djolonga, J.
\newblock You only train once: Loss-conditional training of deep networks.
\newblock In \emph{Proceedings of International conference on learning
  representations}, New Orleans, LSU,USA, 2019.

\bibitem[Goyal et~al.(2019)Goyal, Mahajan, Gupta, and Misra]{goyal2019scaling}
Goyal, P., Mahajan, D., Gupta, A., and Misra, I.
\newblock Scaling and benchmarking self-supervised visual representation
  learning.
\newblock In \emph{Proceedings of the ieee/cvf International Conference on
  computer vision}, pp.\  6391--6400, Seoul, South Korea, 2019.

\bibitem[He et~al.(2016)He, Zhang, Ren, and Sun]{he2016deep}
He, K., Zhang, X., Ren, S., and Sun, J.
\newblock Deep residual learning for image recognition.
\newblock In \emph{Proceedings of the IEEE conference on computer vision and
  pattern recognition}, pp.\  770--778, Las Vegas, NV, USA, 2016.

\bibitem[Kong et~al.(2020)Kong, Cao, Iqbal, Wang, Wang, and
  Plumbley]{Kong19-PANNS}
Kong, Q., Cao, Y., Iqbal, T., Wang, Y., Wang, W., and Plumbley, M.~D.
\newblock Panns: Large-scale pretrained audio neural networks for audio pattern
  recognition.
\newblock \emph{IEEE/ACM Transactions on Audio, Speech, and Language
  Processing}, 28:\penalty0 2880--2894, 2020.

\bibitem[Latif et~al.(2021)Latif, Rana, Khalifa, Jurdak, Qadir, and
  Schuller]{latif2021survey}
Latif, S., Rana, R., Khalifa, S., Jurdak, R., Qadir, J., and Schuller, B.~W.
\newblock Survey of deep representation learning for speech emotion
  recognition.
\newblock \emph{IEEE Transactions on Affective Computing}, pp.\  56--68, 2021.

\bibitem[Liebel \& K{\"o}rner(2018)Liebel and K{\"o}rner]{liebel2018auxiliary}
Liebel, L. and K{\"o}rner, M.
\newblock Auxiliary tasks in multi-task learning.
\newblock \emph{arXiv preprint arXiv:1805.06334}, 2018.

\bibitem[Liu et~al.(2022)Liu, Mallol-Ragolta, Parada-Cabeleiro, Qian, Jing,
  Kathan, Hu, and Schuller]{liu2022audio}
Liu, S., Mallol-Ragolta, A., Parada-Cabeleiro, E., Qian, K., Jing, X., Kathan,
  A., Hu, B., and Schuller, B.~W.
\newblock Audio self-supervised learning: A survey.
\newblock \emph{arXiv preprint arXiv:2203.01205}, 2022.

\bibitem[Mauss \& Robinson(2009)Mauss and Robinson]{mauss2009measures}
Mauss, I.~B. and Robinson, M.~D.
\newblock Measures of emotion: A review.
\newblock \emph{Cognition and emotion}, 23\penalty0 (2):\penalty0 209--237,
  2009.

\bibitem[Park et~al.(2019)Park, Chan, Zhang, Chiu, Zoph, Cubuk, and
  Le]{Park_spec_2019}
Park, D.~S., Chan, W., Zhang, Y., Chiu, C.-C., Zoph, B., Cubuk, E.~D., and Le,
  Q.~V.
\newblock {SpecAugment}: A simple data augmentation method for automatic speech
  recognition.
\newblock \emph{Proceedings of Annual Conference of the International Speech
  Communication Association}, pp.\  2613--2617, 2019.

\bibitem[Scherer et~al.(2017)Scherer, Schuller, and
  Elkins]{scherer2017computational}
Scherer, K.~R., Schuller, B.~W., and Elkins, A.
\newblock Computational analysis of vocal expression of affect: Trends and
  challenges.
\newblock In \emph{{Social Signal Processing}}. Cambridge University Press,
  2017.

\bibitem[Triantafyllopoulos et~al.(2021)Triantafyllopoulos, Liu, and
  Schuller]{Triantafyllopoulos21DSC}
Triantafyllopoulos, A., Liu, S., and Schuller, B.
\newblock {Deep Speaker Conditioning for Speech Emotion Recognition}.
\newblock In \emph{{Proceedings of IEEE International Conference on Multimedia
  and Expo}}, pp.\  1--6, Shenzhen, China, 2021.

\bibitem[Yang et~al.(2020)Yang, Liu, Song, Parada-Cabaleiro, and
  Schuller]{Yang20ARC}
Yang, Z., Liu, S., Song, M., Parada-Cabaleiro, E., and Schuller, B.
\newblock {Adventitious Respiratory Classification using Attentive Residual
  Neural Networks}.
\newblock In \emph{{Proceedings of Annual Conference of the International
  Speech Communication Association}}, pp.\  2912--2916, Shanghai, China, 2020.

\bibitem[Zbontar et~al.(2021)Zbontar, Jing, Misra, LeCun, and
  Deny]{zbontar2021barlow}
Zbontar, J., Jing, L., Misra, I., LeCun, Y., and Deny, S.
\newblock Barlow twins: Self-supervised learning via redundancy reduction.
\newblock In \emph{Proceedings of International Conference on Machine
  Learning}, pp.\  12310--12320, virtual, 2021.

\end{thebibliography}
\bibliographystyle{icml2022}
\end{document}